\documentclass[a4paper,11pt]{article}
\usepackage[utf8]{inputenc}
\usepackage{fullpage}
\usepackage{amsmath,amsthm,amssymb,amsfonts,mathrsfs}
\usepackage{dsfont}
\usepackage{graphicx}
\usepackage{enumitem}
\usepackage{framed}
\usepackage{authblk}
\usepackage{color}

\usepackage{hyperref}
\hypersetup{
	colorlinks = true,
	linkcolor = blue,
	citecolor = blue,
	urlcolor = blue,
	filecolor = black,
	linktoc = all,
}

\usepackage{cite}

\newcommand{\tr}[1]{\operatorname{Tr}\!\left[#1\right]}

\def\>{\rangle}
\def\<{\langle}

\def\id{\mathsf{id}}
\def\mE{\mathcal{E}}
\def\mES{\widetilde{\mathcal{E}}_{\sigma}}
\def\mMS{\widetilde{\mathcal{M}}_{\openone/d}}

\def\mN{\mathcal{N}}
\def\mL{\mathcal{L}}

\def\sH{\mathcal{H}}
\def\sK{\mathcal{K}}

\def\openone{\mathds{1}}

\newcommand{\Supp}{\operatorname{supp}}

\newcommand{\Tr}{\operatorname{Tr}}
\newcommand{\M}{\mathcal{M}}
\newcommand{\one}{\openone}
\newcommand{\tothetimesn}{^{\otimes n}}
\newcommand{\bp}{\boldsymbol{p}}
\newcommand{\bq}{\boldsymbol{q}}
\newcommand{\Dkl}{D_{\operatorname{KL}}}

\newcommand{\Rrec}{\rho_{\operatorname{cg}}}
\renewcommand{\ge}{\geqslant}
\renewcommand{\le}{\leqslant}

\newcommand{\A}{\mathcal{A}}
\newcommand{\B}{\mathcal{B}}
\newcommand{\D}{\mathcal{D}}

\newcommand{\mM}{\mathcal{M}}
\newcommand{\C}{{\mathcal{C}}}

\newcommand{\bra}[1]{\langle #1 |}
\newcommand{\ket}[1]{|#1\rangle }
\newcommand{\ketbra}[1]{|#1\rangle \langle #1 |}

\newcommand{\R}{{\rho}}
\renewcommand{\S}{{\sigma}}
\renewcommand{\[}{\begin{equation}}
\renewcommand{\]}{\end{equation}}

\newcommand{\DKL}{D_{\operatorname{KL}}}

\usepackage{soul,color}

\newtheorem{theorem}{Theorem} 

\newtheorem{corollary}{Corollary}[theorem]
\newtheorem{definition}{Definition}

\theoremstyle{definition}

\theoremstyle{remark}
\newtheorem{remark}{Remark}

\begin{document}

\title{
Observational entropy, coarse-grained states, and the Petz recovery map: information-theoretic properties and bounds
}

\renewcommand\Affilfont{\small \it}

\author[1]{Francesco Buscemi\thanks{buscemi@i.nagoya-u.ac.jp}}
\author[2]{Joseph Schindler\thanks{josephc.schindler@uab.cat}}
\author[3]{Dominik \v{S}afr\'{a}nek\thanks{dsafranekibs@gmail.com}}
\affil[1]{Department of Mathematical Informatics, Nagoya University, Furo-cho, Chikusa-ku, 464-8601 Japan}
\affil[2]{F\'{\i}sica Te\`{o}rica: Informaci\'{o} i Fen\`{o}mens Qu\`{a}ntics, Departament de F\'{\i}sica, Universitat~Aut\`{o}noma~de~Barcelona, 08193 Bellaterra, Spain}
\affil[3]{Center for Theoretical Physics of Complex Systems, Institute for Basic Science (IBS), Daejeon 34126, Republic of Korea}

\date{\today}

\maketitle

\begin{abstract}
    Observational entropy provides a general notion of quantum entropy that appropriately interpolates between Boltzmann's and Gibbs' entropies, and has recently been argued to provide a useful measure of out-of-equilibrium thermodynamic entropy. Here we study the mathematical properties of observational entropy from an information-theoretic viewpoint, making use of recently strengthened forms of the monotonicity property of quantum relative entropy. We present new bounds on observational entropy applying in general, as well as bounds and identities related to sequential and post-processed measurements. A central role in this work is played by what we call the ``coarse-grained'' state, which emerges from the measurement's statistics by Bayesian retrodiction, without presuming any knowledge about the ``true'' underlying state being measured. The degree of distinguishability between such a coarse-grained state and the true (but generally unobservable) one is shown to provide upper and lower bounds on the difference between observational and von Neumann entropies.
\end{abstract}


\section{Introduction}

The idea that coarse-graining plays an important role in statistical thermodynamics is one with a long history---appearing already in early works on statistical mechanics~\cite{gibbs2010elementary,ehrenfest1907begriffliche} (see also~\cite{wehrl-1978-general-entropy}), and first formalized in the context of quantum systems by von Neumann in 1929, within a seminal paper on the quantum Boltzmann H-theorem~\cite{vonNeumann1929translation}.  Though von Neumann there argued for the use of coarse-grained entropy in analyzing thermodynamic systems%
\footnote{Stating, e.g., in the English translation~\cite{vonNeumann1929translation}: ``The expressions for entropy given by the author [previously] are not applicable here in the way they were intended, as they were computed from the perspective of an observer who can carry out all measurements that are possible in principle – i.e., regardless of whether they are macroscopic (for example, there every pure state has entropy 0, only mixtures have entropies greater than 0!). If we take into account that the observer can measure only macroscopically then we find different entropy values (in fact, greater ones, as the observer is now less skilful and possibly can therefore extract less mechanical work from the system).''}, and also devoted an entire section (titled ``the macroscopic measurement'') to this in his book~\cite{von1955mathematical},
for a long time his definition of this coarse-grained entropy was largely forgotten---appearing only sporadically in the literature, and overshadowed by von Neumann's other (more famous and, perhaps~somewhat ironically, eponymous) entropy.

Recent advances, however, have sparked a resurgence of interest in coarse-grained entropies, following the introduction by \v{S}afr\'{a}nek, Deutsch, and Aguirre~\cite{safranek2019b,safranek2019a,safranek2021brief} of ``observational entropy'' as a general framework extending von Neumann's original definition to multiple projective coarse-grainings. Later, also general coarse-grainings (quantum instruments) were included~\cite{safranex2021generalized}. This framework has the benefit that a ``coarse-graining'' $\mathcal{C}$ may be defined by any quantum measurement (any quantum instrument or POVM), with different choices of coarse-graining relevant to different physical or experimental scenarios. Associated to a given coarse-graining and a state~$\rho$, then, is the observational entropy $S_\mathcal{C}(\rho)$, a measure of uncertainty associated to~$\rho$ under measurement~$\mathcal{C}$, which includes contributions from both Gibbs-like and Boltzmann-like entropy terms. 

Traditional thermodynamic entropies arise, in this context, by considering suitably chosen coarse-grainings, and a number of recent studies support the idea that observational entropy is indeed an appropriate quantity to describe statistical thermodynamics in both equilibrium and non-equilibrium systems, much along the lines of the original von Neumann's reasoning. Included among these are applications to thermalization in isolated systems~%
\cite{safranek2019b,safranek2019a,safranek2021brief,safranex2021generalized}, 
heat transfer and entropy production in open systems coupled to a bath~%
\cite{strasberg2020first,riera2020finite}, 
comparison of classical and quantum entropies~%
\cite{safranek2020classical}, 
and further applications in quantum statistical thermodynamics~%
\mbox{\cite{deutsch2018eth,deutsch2020probabilistic,faiez2020typical,nation2020snapshots,strasberg2021clausius,hamazaki2022speed,zhou2022renyi}}, work extraction from unknown sources~\cite{safranek2022work},
entanglement/correlation theory~%
\cite{schindler2020correlation,zhou2022correlation},
and quantum cosmology~%
\cite{amadei2019unitarity,lun2020vacuum}.
We also mention other works that can be rephrased in the context of observational entropy~\cite{tolman1938principles,ter1954elements,jaynes1957information2,ingarden1962quantum,grabowski1977continuity,wehrl1979relation,lieb1979proof,polkovnikov2011microscopic,santos2011entropy,giraud2016average,anza2017information,anza2018new,anza2018eigenstate,lent2019quantum}.

Motivated by these exciting developments, in this paper we turn to the consideration of observational entropy from an information-theoretic perspective---focusing on the consequences of relative entropy monotonicity under completely positive trace-preserving (CPTP) maps, where recent advances can provide finite bounds on entropy difference in the case of approximate recoverability~\cite{petz1986sufficient,petz1988sufficiency,fawzi-renner-2015,wilde2015recoverability,seshadreesan-wilde-2015-fid-recov,brandao-etal-2015-recostructed-states,sutter-tomamichel-harrow,buscemi-das-wilde,wilde2017quantum-book,li-winter,junge,sutter2017multivariate}. 

The program is a simple one. First we encode the coarse-graining $\mathcal{C}$ in a quantum-classical channel $\mathcal{M}$ (see Eq.~\eqref{eq:M} below), allowing observational entropy to be related to a quantum relative entropy $D(\mathcal{M}(\rho)||\mathcal{M}(\openone/d))$. The measurement map can be reversed in the sense of Petz recovery. The recovered state $\Rrec$ is shown to take on a simple form in terms of the POVM elements~$\Pi_i$ and probabilities $p_i=\tr{\Pi_i\ \rho}$ associated with the coarse-graining, being given by
\begin{equation}
    \Rrec = \sum_i p_i \; \Pi_i / \tr{\Pi_i}.
\end{equation}
This state is shown to correspond to a Bayesian estimate of $\rho$ given only coarse-grained knowledge, and is seen to arise from both the Petz~\cite{petz1986sufficient,petz1988sufficiency} and the ``rotated'' Petz~\cite{junge} recovery operations. Thus, in terms of known bounds on relative entropy difference, we are able to obtain new bounds on observational entropy $S_{\C}(\rho)$, including in particular the bound (with~$S(\rho)$ von Neumann entropy and $D$ quantum relative entropy)
\begin{equation}
\label{eqn:intro-Fisher-bound}
    S_{\C}(\rho) - S(\rho) \ge D(\R\|\Rrec)\;.
\end{equation}
Thus observational entropy provides a quantitative measure of closeness between the ``true'' state~$\rho$ and the inferred coarse state~$\Rrec$, elevating the status of the coarse-grained state from a statistical estimate to a practical approximation. A converse bound will also be shown.

In addition to demonstrating the bound above, we also investigate observational entropy associated with other scenarios of interest, such as sequential measurements, refined measurements, and convex combinations of measurements, and analyze these in terms of Petz recovery.

After briefly introducing basic notions and notations in Section~\ref{sec:definition}, in Section~\ref{sec:approx} we define the coarse-grained state and obtain the main recovery inequality \eqref{eqn:intro-Fisher-bound} and its converse as Theorem~\ref{thm:approx-lower-bound}. In section~\ref{sec:bayesian} we discuss the interpretation of the recovery inequality and coarse-grained state in terms of Bayesian retrodiction. Sections~\ref{sec:sequential} and \ref{sec:refine}  consider observational entropy under sequential measurements and refinements of measurements. Section~\ref{sec:concave} discusses concavity over states and measurements. Bounds and exact chain rules for sequential measurement appear in Theorems~\ref{thm:approx-non-increase},\ref{thm:sandwich} and for refinements in Theorem~\ref{thm:approx-monotonic2}.

\section{Definitions and notation}\label{sec:definition}

In this work we will deal exclusively with finite quantum systems, i.e., systems associated to finite dimensional Hilbert spaces, denoted as $\sH$, $\sK$, etc. The dimension of the space will be denoted by $d$. In this case, quantum states are represented by density matrices, i.e, positive semidefinite linear operators $\rho\ge 0$ with unit trace. The support of $\rho$ is defined as the orthogonal complement to its kernel, and is denoted by $\Supp(\rho)$. Given two quantum states, $\rho$ and $\sigma$, defined on the same Hilbert space, it is possible to measure their closeness in various ways. One such way is given by the \textit{trace distance}, defined as $T(\rho,\sigma):=\frac12\tr{\;|\rho-\sigma|\;}\equiv\frac{1}{2}\left\|\rho-\sigma\right\|_1$, which is equal to one if and only if the two states have orthogonal supports, and zero if and only if $\rho=\sigma$. The trace distance is directly related to the probability of correctly distinguishing between $\rho$ and $\sigma$, a fundamental result known as Helstrom bound~\cite{helstrom-1969}. Another widely used measure of closeness between two quantum states is the \textit{quantum fidelity}~\cite{UHLMANN1976-fidelity,jozsa-1994-fidelity}, which is defined as $F(\rho,\sigma):=\Tr\left[\sqrt{\sqrt{\sigma}\rho\sqrt{\sigma}}\right]$. The fidelity is related to the trace distance by $1-F(\rho,\sigma)\le T(\rho,\sigma)\le \sqrt{1-\left[F(\rho,\sigma)\right]^2}$~\cite{fuchs-vandegraaf-1999-bounds-distances}. A third, very important, measure of statistical distinguishability between quantum states is given by the \textit{Umegaki quantum relative entropy}, which for two states $\rho$ and $\sigma$ is defined as~\cite{umegaki-q-rel-ent-1961,umegaki1962conditional}
\[\label{eq:umegaki}
D(\R\|\S):=
\begin{cases}
\tr{\R\ln\R}-\tr{\R\ln\S}\;,& \text{if }\Supp(\R)\subset \Supp(\S)\;,\\
+\infty\;, & \text{otherwise}\;.
\end{cases}
\]

\begin{remark}\label{rem:KL-divergence}
The Umegaki quantum relative entropy $D$ was introduced as a generalization of Kullback--Leibler divergence $\DKL$~\cite{kullback1951}, often referred to as \emph{classical} relative entropy.
\end{remark}\bigskip 

The quantum relative entropy $D(\rho\|\sigma)$ is related to the trace distance $T(\rho,\sigma)$ and the fidelity $F(\rho,\sigma)$ via two bounds: the quantum Pinsker inequality, i.e., $D(\rho\|\sigma)\ge 2[T(\rho,\sigma)]^2$, see~\cite{ohya2004quantum}, and  $D(\rho\|\sigma)\ge -2\ln F(\rho,\sigma)$, see, e.g., Eq.~(5.43) in~\cite{hayashi2006quantum}. 

The von Neumann entropy of a state $\rho$ is defined as~\cite{von1955mathematical}
\[\label{eq:usual-entropy}
S(\R):=-\tr{\R\ln\R}\;.
\]
It is a well-known fact that von Neumann's entropy can be expressed in terms of Umegaki's relative entropy as follows (see, e.g., Ref.~\cite{wilde2017quantum-book}):
\[\label{eq:E_as_relativeE}
S(\R)=-D(\rho\|\openone)=\ln d-D(\rho\|u)\;,
\]
where $\openone$ denotes the identity matrix and $u$ is a shorthand notation for the maximally mixed state $\openone/d$.

When dealing with open quantum systems, a central notion is that of \textit{quantum operations}~\cite{kraus1983states}, i.e., completely positive (CP) linear maps transforming quantum states on one Hilbert space to another. Any quantum operation $\A$ mapping operators on $\sH$ to operators on $\sK$ admits the following (Kraus) representation~\cite{choi1975completely,kraus1983states}:
\begin{align*}
    \A(\bullet)=\sum_{k=1}^N A_k\bullet A_k^\dag\;,
\end{align*}
where $\bullet$ denotes a generic input (i.e., a matrix on $\sH$), $A_k$ are linear operators from $\sH$ to $\sK$, and the superscript $\dag$ represents the complex conjugate transpose (i.e., the dagger involution). The operators $A_k$ are called the \textit{Kraus operators} associated to the operation $\A$. A quantum operation is called a \textit{quantum channel} whenever it is trace-preserving (TP), that is, $\Tr[\A(X)]=\Tr[X]$, for all operators $X$. Therefore, in what follows the terms channel and CPTP (i.e., completely positive trace-preserving) linear map will be used interchangeably.

The TP condition can be equivalently expressed in terms of the Kraus operators as follows
\begin{align*}
    \Tr[\A(X)]=\Tr[X]\iff\sum_{k=1}^N A_k^\dag A_k=\openone_{\sH}\;.
\end{align*}
By defining the trace-dual linear map $\A^\dag$ as the unique linear map such that $\Tr[Y\ \A(X)]=\Tr[\A^\dag(Y) X]$, for all $X$ on $\sH$ and all $Y$ on $\sK$, we have that
\begin{align}\label{eq:trace-dual}
    \A^\dag(\bullet)=\sum_{k=1}^N A^\dag_k\bullet A_k\;,
\end{align}
where now the input $\bullet$ is a matrix on $\sK$, that is, the linear map $\A^\dag$ takes linear operators on $\sK$ to linear operators on $\sH$. Notice that the quantum operation $\A$ is trace-preserving if and only if $\A^\dag(\openone_\sK)=\openone_\sH$.

In this paper, by \textit{coarse-graining} we denote a quantum instrument~\cite{davies1970operational,ozawa1984quantum}, namely, a finite collection $\C$ of quantum operations $\{\A_i \}_{i}$ labeled by an index $i\in I$, such that $\sum_{i\in I}\A_i$ is trace-preserving. Quantum instruments are used to model quantum measurement processes: given an initial quantum state $\rho$, the probability of obtaining the outcome $i$ is given by $p_i=\Tr[\A_i(\rho)]$, while the corresponding post-measurement state is $\A_i(\rho)/p_i$. When only the outcome distribution is of interest, it is not necessary to work with the operations $\A_i$: it suffices to consider the \textit{positive operator-valued measure} (POVM) associated to $\C$, namely, the family of positive semidefinite operators $\Pi_i$ defined by (see Eq.~\eqref{eq:trace-dual} again) $\Pi_i=\A_i^\dag(\openone)$, so that
\begin{align*}
p_i=\Tr[\A_i(\rho)]=\Tr[\A^\dag_i(\openone)\ \rho]=\Tr[\Pi_i\ \rho]\;.
\end{align*}
It is immediate to verify that the condition that the average transformation $\sum_i\A_i$ is trace-preserving is equivalent to the condition that $\sum_i\Pi_i=\openone$. We will call a coarse-graining \textit{projective} if all its POVM elements are mutually orthogonal projectors.

\begin{figure}[t!]
	\centering
\includegraphics[width=0.5\linewidth]{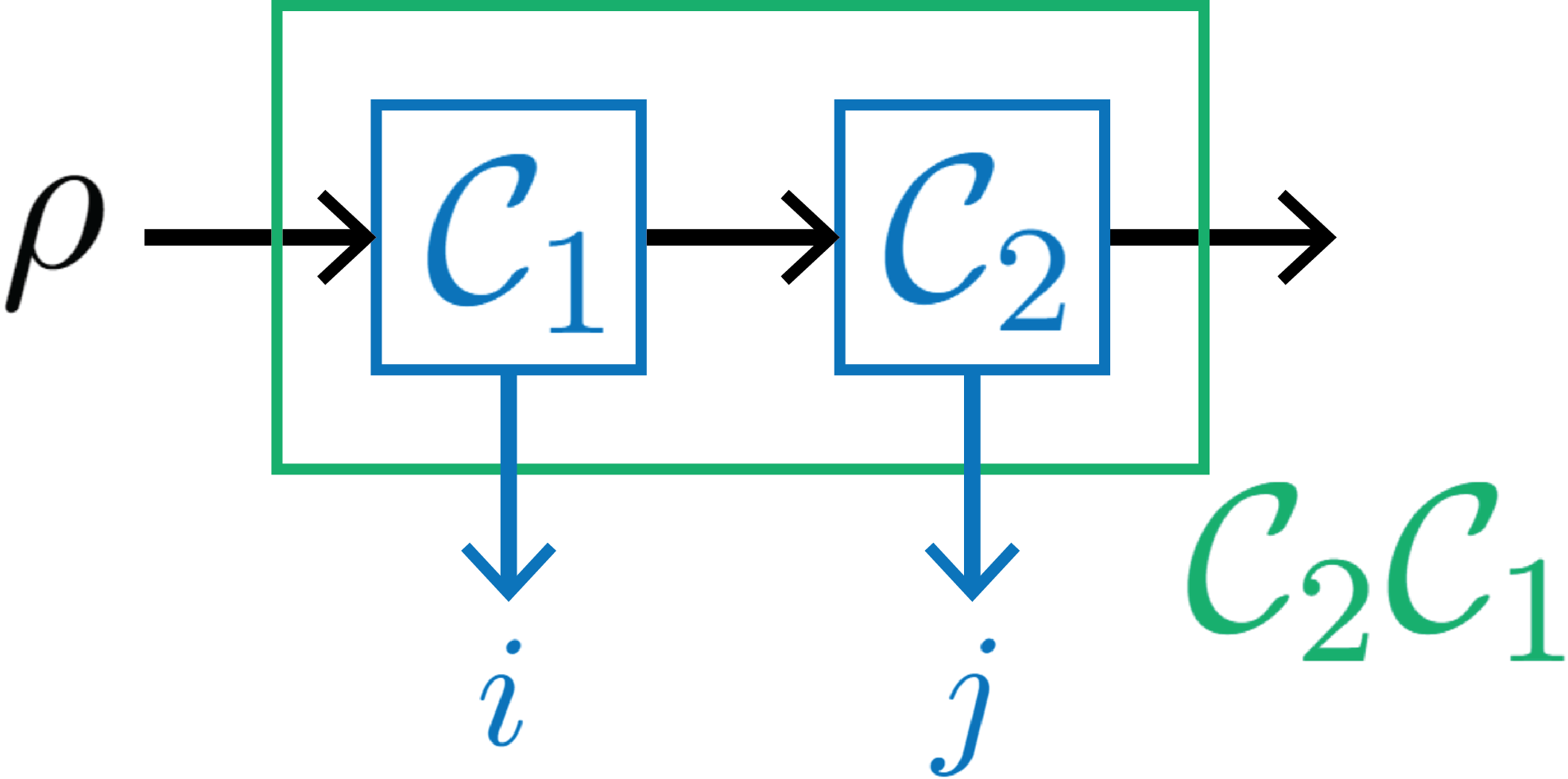}
	\caption{
	Given two coarse-grainings $\C_1=\{\A_i\}_{i\in I}$  and $\C_2=\{\B_j\}_{j\in J}$, whenever the output system of $\C_1$ coincides with the input system of $\C_2$, the two coarse-grainings are composable, in the sense that it is possible to consider their composition in series, denoted by $\C_2\C_1$, which corresponds to the family of quantum operations $\{\B_j\circ\A_i\}_{i,j}$ labeled by the double index $(i,j)\in I\times J$, with the composition symbol $\circ$ to be read as ``after''. Notice that the notation $\C_2\C_1$ follows the convention of composition of operators, that is, $\C_1$ is applied first. The same construction can be straightforwardly extended to sequences of $n$ composable coarse-grainings.
	}
	\label{fig:coarse-graining}
\end{figure}

The main reason to consider quantum instruments, instead of POVMs, is that they can, differently from POVMs, be arranged in sequential series, as shown in Fig.~\ref{fig:coarse-graining}. Consider two coarse-grainings $\C_1 =\{\A_i\}$ and $\C_2=\{\B_j\}$, and assume that they are composable, that is, the output system of $\C_1$ coincides with the input system of $\C_2$. We can then consider the situation in which we measure in series $\C_1$ first and then $\C_2$. The resulting sequence, which we denote for brevity as $\C_2\C_1$, algebraically corresponds to applying in series $\C_1$ followed by $\C_2$. We thus obtain that the joint probability of obtaining outcome $i$ and then outcome $j$ is given by 
\[\label{eq:pi}
\begin{split}
 	p_{ij}&\equiv p(i,j)\\
 	&=\Tr[(\B_{j}\circ \A_{i})(\rho) ]\\
 	&=\Tr[(\A_i^\dag\circ\B_j^\dag)(\openone)\ \rho]\\
	&=\Tr[\Pi_{ij}\, \rho]\;,   
\end{split}
\]
where the symbol $\circ$ denotes the composition in series (to be read ``after''), e.g., $(\B_j\circ\A_i)(\rho)=\B_j(\A_i(\rho))$. From the above, it becomes clear that a sequence of coarse-grainings corresponds to a single coarse-graining with two outcomes, defined as $\C_2\C_1=\{\B_j\circ\A_i\}_{i,j}$. When needed, sequences of multiple coarse-grainings will be denoted as $\C_n\cdots\C_1$, which corresponds to $n$ measurements performed sequentially.

\section{Observational entropy's recovery inequality}
\label{sec:approx}

The main quantity considered in this work is the following:
\begin{definition}[Observational entropy~\cite{vonNeumann1929translation,von1955mathematical,safranek2019b,safranex2021generalized}]
    The observational entropy of a state $\R$ with respect to a coarse-graining $\C=\{\A_i\}_i$ is defined as
\[\label{eq:obsE}
S_{\C}(\R):=-\sum_{i\in I} p_i\ln \frac{p_i}{V_i}\;,
\]
where $p_i$ and $V_i$ are the ``probability terms'' and ``volume terms'' given by
\begin{align}\label{eq:pi-vi}
    p_i:=\tr{\A_i(\R)}   = \tr{\Pi_i \rho}\;,\qquad
    V_i:= \tr{\Pi_i}\;,
\end{align}
respectively.
\end{definition}
The above definition applied to pairs of coarse-grainings $\C_1=\{\A_i\}_{i}$  and $\C_2=\{\B_j\}_{j}$ measured sequentially as in Fig.~\ref{fig:coarse-graining} gives $S_{\C_2\C_1}(\R) = -\sum_{i,j} p_{ij}\ln (p_{ij}/V_{ij})$, where
\begin{align}\label{eq:pi-vi-two}
    p_{ij}=\tr{\B_j(\A_i(\R))}   = \tr{\Pi_{ij} \rho}\;,\qquad
    V_{ij}= \tr{\Pi_{ij}}\;.
\end{align}

As it happens for von Neumann's entropy~\eqref{eq:E_as_relativeE}, the observational entropy too can be written in terms of the quantum relative entropy. By associating to each coarse-graining $\C=\{\A_i\}_i$ a CPTP map $\mM$ that maps its outcomes onto distinguishable (i.e., orthogonal) pure states of the measurement device, namely,
\[\label{eq:M}
\mM(\bullet)=\sum_{i\in I}\tr{\Pi_i\ \bullet}\ketbra{i}\;,
\]
we can write the observational entropy~\eqref{eq:obsE} as
\[\label{eq:obsE_as_relativeE}
S_{\C}(\R)=-D(\mM(\R)\|\mM(\openone))=\ln d-D(\mM(\R)\|\mM(u))\;,
\]
where, we recall, $u:=\openone/d$.
Channels like $\mM$ are often referred to in the literature as ``quantum-to-classical'' or qc-channels.

By combining Eqs.~\eqref{eq:obsE_as_relativeE} and~\eqref{eq:E_as_relativeE} we immediately see that
\begin{align}\label{eq:diff-ent-diff-rel}
S_{\C}(\R)-S(\R)=D(\R\|u)-D(\mM(\R)\|\mM(u))\;,
\end{align}
that is, the difference between observational entropy and von Neumann entropy equals the decrease in relative entropy due to the action of the qc-channel $\mM$ defined in~\eqref{eq:M}, between the given state $\R$ and the maximally mixed (uniform) state $u=\openone/d$. This simple observation provides a direct link between observational entropy and a recently very active area of quantum information theory known collectively as ``approximate recoverability theory''~\cite{fawzi-renner-2015,wilde2015recoverability,seshadreesan-wilde-2015-fid-recov,brandao-etal-2015-recostructed-states,sutter-tomamichel-harrow,buscemi-das-wilde,wilde2017quantum-book,li-winter,junge,sutter2017multivariate}, which has found various applications also in other areas of mathematical physics~\cite{entanglement-wedge-2019}. In what follows, we will use this connection as the starting point to derive a number of new results about observational entropy.

Before doing that, however, we begin this section from the most fundamental property of quantum relative entropy, that is, its monotonicity property under the action of channels. This property, together with the conditions for equality, is summarized by Petz's famous recovery theorem (of which the theory of approximate recoverability is a generalization).

\begin{theorem}\label{thm:inequality_for_channels}
For all channels $\mE$ and all states $\R$, $\S$, we have
\[\label{eq:PetzInequality}
D(\R\|\S)\ge D(\mE(\R)\|\mE(\S))\;,
\]
with equality if and only if the CPTP map defined as
\[\label{eq:Petz_map}
\mES(\bullet):=\sqrt{\S}\mE^\dag\left[\frac{1}{\sqrt{\mE(\S)}}\bullet\frac{1}{\sqrt{\mE(\S)}}\right]\sqrt{\S}\;,
\]
where $\mE^\dagger$ is defined by trace-duality as in Eq.~\eqref{eq:trace-dual},
satisfies
\[
\mES(\mE(\R))=\R\;.
\]
(The other equality, i.e., $\mES(\mE(\S))=\S$, is satisfied by construction.)
\end{theorem}

The inequality in Theorem~\ref{thm:inequality_for_channels} was proved in the 1970s by Lindblad~\cite{lindblad1975completely} and Uhlmann~\cite{uhlmann1977relative}, while the condition for equality was proved by Petz~\cite{petz1986sufficient,petz1988sufficiency} a few years later (see also~\cite{hayden2004structure} for a pedagogical approach). For this reason, the CPTP map defined in~\eqref{eq:Petz_map} is often referred to as the ``Petz recovery map''. From Theorem~\ref{thm:inequality_for_channels}, various properties of the observational entropy follow as corollaries.

Applying the Petz recovery map to the case of a measurement's qc-channel, relative to the uniform reference state, results in a ``coarse-grained'' state $\Rrec$, which plays a central role in the treatment of observational entropy to follow.

\begin{definition}[Coarse-grained state]
For any coarse-graining $\C$ with associated POVM $\{\Pi_i\}_{i\in I}$, and any quantum state $\rho$, the corresponding \emph{coarse-grained state} is defined by
	\begin{align}\label{eq:recovered-state}
	    \Rrec :=\sum_{i\in I} \frac{p_i}{V_i} \, \Pi_i\;,
	\end{align}
with $p_i$ and $V_i$ given in Eq.~\eqref{eq:pi-vi}.
\end{definition}

\begin{remark}   The coarse-grained state already appears in Wehrl~\cite[(1.37)]{wehrl-1978-general-entropy}. Note, however, that Wehrl worked in the context of projective measurements (i.e., $\Pi_i\Pi_j=\delta_{i,j}\Pi_i$), whereas Eq.~\eqref{eq:recovered-state} is given for general coarse-grainings associated with possibly non-projective POVMs.
\end{remark}\bigskip

As a first step, we recover the lower bound on observational entropy framed concisely in terms of the coarse-grained state.

\begin{theorem}\label{th:lower-bound-vN-E}
Let $\C$ be a coarse-graining and $\{\Pi_i\}_i$ its associated POVM. For any state $\R$
\[\label{eq:simple-bound}
S_{\C}(\R)\ge S(\R)\;,
\]
with equality if and only if
\[\label{eq:retro-state}
\R=\Rrec\;.
\]
\end{theorem}

An equivalent form of this theorem appeared previously in~\cite{safranex2021generalized}, but in Appendix~\ref{sec:app-proof-of-bound-with-vonN-entropy} we show that it follows directly from Theorem~\ref{thm:inequality_for_channels} applied to the qc-channel $\mM$, together with the observation that the Petz recovery map in this case takes on the simple form
\begin{align}
\widetilde{\mM}_u(\mM(\bullet))=\sum_i\frac{\tr{\Pi_i\ \bullet}}{V_i} \, \Pi_i\;.
\end{align}

\begin{remark}
    Notice that normalizing each POVM element $\Pi_i$ by its trace, we obtain a family of density matrices $\Pi_i/V_i$, and Eq.~\eqref{eq:retro-state} expresses the fact that these provide a convex decomposition for all those $\rho$ such that $S_{\C}(\rho)=S(\rho)$.
\end{remark}
\vspace{6pt}

The strengthening of Theorem~\ref{thm:inequality_for_channels} has been one of the most important developments in quantum information theory in recent years. In this section, we investigate what the theory of approximate recoverability~\cite{fawzi-renner-2015,wilde2015recoverability,seshadreesan-wilde-2015-fid-recov,brandao-etal-2015-recostructed-states,sutter-tomamichel-harrow,buscemi-das-wilde,wilde2017quantum-book,li-winter,junge,sutter2017multivariate} can tell us about the observational entropy. We begin with the following result, which extends the scope of Theorem~\ref{th:lower-bound-vN-E} in various ways.

\begin{framed}
\begin{theorem}[Observational entropy's recovery inequality]
\label{thm:approx-lower-bound}
For any coarse-graining $\C$, with associated POVM $\{\Pi_i\}_{i\in I}$, we have
\begin{align}
S_\C(\rho)-S(\rho)\ge { D\left(\R\middle\|\Rrec\right)\;.}\label{eq:quantum_rel_entropy_bound}
\end{align}
Conversely,
\begin{align}\label{eq:upper-bound-fannes}
   S_\C(\rho)-S(\rho)\le T(\rho,\Rrec)\ln(d-1)+h[T(\rho,\Rrec)]\;,
\end{align}
where $h(x):=-x\ln x-(1-x)\ln(1-x)$ is the binary entropy and $T(\rho,\sigma):=\frac12\tr{\;|\rho-\sigma|\;}$ is the trace distance (see Sec.~\ref{sec:definition}).
\end{theorem}
\end{framed}

\begin{remark}
   The above theorem implies Theorem~\ref{th:lower-bound-vN-E} as a corollary: the inequality in Theorem~\ref{th:lower-bound-vN-E} comes from the non-negativity of the quantum relative entropy, while the equality condition arises from the fact that the right-hand side in~\eqref{eq:upper-bound-fannes} smoothly goes to zero as the trace distance $T(\rho,\Rrec)\to 0$.
\end{remark}
\vspace{6pt}

\begin{remark}
	An important point to stress is that $\Rrec$ encodes all the information that is available to the experimenter. This is so, because $p_i$ can be estimated from the statistics of occurrence of each outcome, while the POVM elements $\{\Pi_i\}$ describe the experimental setup. The theorem says that the closer the observational entropy (which also depends solely on the $p_i$'s and the $\Pi_i$'s) is to the von Neumann entropy, the closer such a coarse-grained state is to the state of the system undergoing the measurement. Unfortunately, one cannot assume that the ``true'' von Neumann entropy is always known, so that a comparison with the observational entropy can always be done. However, it is a remarkable consequence of Theorem~\ref{thm:approx-lower-bound} that the sole knowledge of the initial ``true'' entropy (a real number) allows for an estimation of the entire ``true'' state (a density matrix).
\end{remark}
\vspace{6pt}

\begin{remark}
The bound~\eqref{eq:quantum_rel_entropy_bound} above implies a triangle-like inequality between the quantum state~$\R$, its Petz recovered version $\Rrec$, and the maximally mixed state $u$, as follows:
\[
D(\rho\|u)\ge D(\R\|\Rrec)+D(\Rrec\|u)\;.
\]
The above relation can be shown by noticing that~\eqref{eq:quantum_rel_entropy_bound} can be rewritten as
\begin{equation*}
    D(\rho\|u)\ge D(\R\|\Rrec)+D(\mM(\R)\|\mM(u))\;,
\end{equation*}
while $D(\mM(\R)\|\mM(u))\ge D(\widetilde{\mM}_u(\mM(\R))\|\widetilde{\mM}_u(\mM(u)))= D(\Rrec\|u)$.
\end{remark}
\vspace{6pt} 

\begin{remark}
The bound~\eqref{eq:quantum_rel_entropy_bound} also implies a bound in terms of quantum fidelity and trace distance: see the relations between these and the quantum relative entropy below Eq.~\eqref{eq:umegaki}.
\end{remark}
\vspace{6pt}

\begin{proof}[Proof of Theorem~\ref{thm:approx-lower-bound}]
We begin with the proof of bound~\eqref{eq:quantum_rel_entropy_bound}. Making use of the strong form of relative entropy monotonicity given by Corollary 4.2 of~\cite{sutter2017multivariate} it follows that for arbitrary state $\rho$ and coarse-graining $\C$,
\begin{equation}
\label{eq:sutter-bound}
    S_\C(\rho) - S(\rho) = D(\rho\|u) - D(\M(\rho)\|\M(u)) \ge D_{\mathbb{M}}(\rho \| R),
\end{equation}
where $R$ is a state obtained from $\mM(\R)$ by acting upon it with a suitable ``approximate recovery channel'' $\mathcal{R}_{u,\mM}$, that is, $R:=\mathcal{R}_{u,\mM}(\mM(\R))$, cf. Eq.~(61) of~\cite{sutter2017multivariate}, while $D_{\mathbb{M}}$ is the ``measured relative entropy'', i.e., the supremum over all POVMs of the classical relative entropy of the outcome statistics, cf.~\cite{sutter2017multivariate,junge,hiai1991proper}. In general $D_\mathbb{M}(\rho \| R) \le D(\rho \| R)$, the state $R$ may be different from the coarse-grained state $\Rrec$, and the bound \eqref{eq:sutter-bound} cannot be strengthened further. In the case relevant here, however, the reference prior is the uniform state $u$, which commutes with any other operators, and $\M$ is a quantum-classical map, whose outputs all commute with each other. These two features together allow several simplifications, ultimately leading to~\eqref{eq:quantum_rel_entropy_bound}.

To show how this is done, we consider an asymptotic $n$-copy limit. Consider applying the measurement $\C' =\C\tothetimesn \equiv (\C_{i_1} \otimes \ldots \otimes \C_{i_n})_{i_1, \ldots, i_n}$ to the state $\rho' =\rho\tothetimesn$. The associated measuring channel to $\C'$ is $\M'(\bullet) = \sum_{i_1, \ldots, i_n} \tr{\Pi_{i_1} 
\otimes \ldots \otimes \Pi_{i_n}\ \bullet} \ketbra{i_1, \ldots, i_n} = \M\tothetimesn(\bullet)$, which follows from linearity with equality on general product states. We therefore have
\begin{align}
    S_{\C'}(\rho') - S(\rho') &= D\left(\rho\tothetimesn\left\|u\tothetimesn\right)\right. - D\left(\M\tothetimesn[\rho\tothetimesn]\left\|\M\tothetimesn[u\tothetimesn]\right.\right) \\
    &\ge D_{\mathbb{M}}(\rho\tothetimesn \| R'),
\end{align}
where the recovered state is defined as $R' = (\mathcal{R}_{u\tothetimesn, \M\tothetimesn} \circ \M\tothetimesn)(\rho\tothetimesn)$.

We can now find that in this special case $R' = \Rrec\tothetimesn$ (although the same does not hold in general). Consider Eq.~(61) of \cite{sutter2017multivariate} which defines the rotated Petz recovery map for a CPTP map $\M$ and state $\sigma$ as,
\[
\mathcal{R}_{\sigma,\M}:=\int_{-\infty}^{\infty}dt\, \beta_0(t)\mathcal{R}_{\sigma,\M}^{[t]}(\bullet)\quad\text{and}\quad
\mathcal{R}_{\sigma,\M}^{[t]}:=\sigma^{\frac{1+it}{2}}\mathcal{M}^\dag\Big(\mathcal{M}(\sigma)^{-\frac{1+it}{2}}(\bullet)\mathcal{M}(\sigma)^{-\frac{1-it}{2}}\Big)\sigma^{\frac{1-it}{2}}.
\]
Observe that the state $u\tothetimesn = (\one/d)\tothetimesn$ is a scalar multiple of the identity, and therefore commutes with everything. Further, observe that because $\M'(\bullet)$ outputs classical states (diagonal in a fixed basis), it follows that $\M'(\rho')$ and $\M'(\sigma')$ mutually commute. Using these two observations it follows straightforwardly that all the rotated Petz maps are equal, in particular
\begin{align}  (\mathcal{R}^{[t]}_{u\tothetimesn, \M\tothetimesn} \circ \M\tothetimesn)(\rho\tothetimesn) &= (\mathcal{R}^{[0]}_{u\tothetimesn, \M\tothetimesn} \circ \M\tothetimesn)(\rho\tothetimesn)\\
    &\equiv (\widetilde{\mM}_u^{\otimes n} \circ \M\tothetimesn)(\rho\tothetimesn)
\end{align}
and therefore further we have
\begin{align}
    R' &= (\widetilde{\mM}_u\circ\mM)^{\otimes n}(\rho^{\otimes n})\\ 
    &= \Rrec^{\otimes n}\;.
\end{align}

On the other hand we have $S_{\C\tothetimesn}(\rho\tothetimesn)=n S_\C(\rho)$, and likewise from the additivity of the von Neumann entropy. Collecting this with the above results one obtains, for all $n$,
\begin{equation}
    n S_{\C}(\rho) - n S(\rho) \ge D_{\mathbb{M}}(\rho\tothetimesn \| \Rrec\tothetimesn).
\end{equation}
It remains only to divide by $n$ and take the $n \to \infty$ limit. Making use of the fact that 
measured relative entropy $\lim_{n \to \infty}  \frac{1}{n} D_\mathbb{M}(\rho\tothetimesn \| \sigma\tothetimesn) = D(\rho \| \sigma)$ asymptotically achieves quantum relative entropy in the many copy limit~\cite{hiai1991proper} (cf. also~\cite{vedral1997statistical,hayashi2001asymptotics}), we have
\begin{equation}
    S_{\C}(\rho) - S(\rho) \ge \lim_{n \to \infty}  \frac{1}{n}  D_{\mathbb{M}}(\rho\tothetimesn \| \Rrec\tothetimesn) = D(\rho \| \Rrec)\;,
\end{equation}
thus establishing the inequality.

We now turn to the converse bound~\eqref{eq:upper-bound-fannes}. By invoking the monotonicity property of the quantum relative entropy twice, that is, $D(\R\|u)\ge D(\mM(\R)\|\mM(u))\ge D(\widetilde{\mM}_u(\mM(\R))\|\widetilde{\mM}_u(\mM(u)))\equiv D(\Rrec\|u)$, we have that
\begin{align*}
    S_\C(\rho)-S(\rho)&=D(\R\|u)- D(\mM(\R)\|\mM(u))\\
    &\le D(\R\|u)- D(\Rrec\|u)\\
    &= S(\Rrec)-S(\R)\\
    &\le T(\rho,\Rrec)\ln(d-1)+h[T(\rho,\Rrec)]\;,
\end{align*}
where for the last inequality we used the Fannes--Audenaert continuity bound for von Neumann entropy~\cite{Fannes,Audenaert}, see also Theorem~11.10.2 in~\cite{wilde2017quantum-book}.
\end{proof}

From Theorem~\ref{thm:approx-lower-bound} and the final passages of its proof we immediately obtain the following, which can be of independent interest:
\begin{corollary}
For any coarse-graining $\C$ and any state $\rho$,
\begin{align}\label{eq:entropy-sandwich}
	S(\Rrec)\ge S_{\C}(\rho)\ge S(\rho)\;.
\end{align}
\end{corollary}
\noindent
When $\C$ is in particular a projective measurement, the left inequality in \eqref{eq:entropy-sandwich} becomes the equality $S(\Rrec) = S_{\C}(\rho)\ge S(\rho)$.

\section{Interpretation of the coarse-grained state as retrodiction}
\label{sec:bayesian}

The statements of Theorems~\ref{th:lower-bound-vN-E} and~\ref{thm:approx-lower-bound} pose the question about the meaning and interpretation of the coarse-grained state
\[\label{eq:retrodiction}
\Rrec=\sum_i\frac{p_i}{V_i}\Pi_i\;.
\]
Does it provide a sort of ``tomographic reconstruction'' of $\rho$? Or is it an ``error corrected'' state? Or something else? In order to answer this question, we begin from the classical setting, where the coarse-grained state enjoys a very clear interpretation in terms of Bayesian inference.

Suppose that an agent has some prior belief about the state $x$ of a (finite, classical) system. Let us represent the agent's belief with a probability distribution $\alpha(x)$. Suppose moreover that the agent later performs an observation on the system: such observation can take values in a finite set $\{y\}$, and the likelihood of each state given a certain outcome (that is, the probability of each outcome given a certain state) is $\mL(y|x)$. Then, Bayes theorem~\cite{jaynes_2003} tells us that, in the face of a particular observation $\bar y$, the agent should update their belief as follows:
\[\nonumber
\alpha(x)\quad\longrightarrow\quad \alpha(x|\bar y)= \frac{\alpha(x)\mL(\bar y|x)}{[\mL\alpha](\bar y)}\;,
\]
where we denoted $[\mL\alpha](y):=\sum_x\alpha(x)\mL(y|x)$.

The connection with Eq.~\eqref{eq:retrodiction} arises not directly from Bayes theorem, but from an extension thereof, which occurs when the agent's observation does not lead to any definite outcome, but to a \textit{further degree of belief} about which outcome actually occurred. The typical example is that of an observation by candlelight\footnote{Suppose we are looking at a piece of cloth in very dim light: we may be able to say with certainty that its color is not, say, yellow, but we may not be able to tell an orange cloth from a brown one with 100\% confidence, hence the updated belief given in the form of a new probability distribution.}. Suppose therefore that the result of such an observation is represented by another probability distribution $\omega(y)$ over all possible outcomes. Notice that $\omega(y)$ can be completely arbitrary: it is just new information that is given to the agent, and it may or may not be consistent with the agent's prior belief. In the literature $\omega(y)$ is often referred to as \textit{soft evidence}, in contrast to the ``hard'' evidence, corresponding to a delta distribution, used in Bayes' theorem. Then, \textit{Jeffrey's rule}~\cite{jeffrey} tells us that the agent should update their belief as follows:
\begin{align*}
	\alpha(x)\quad\longrightarrow\quad \alpha(x|\omega)= \sum_y\tilde\mL_\alpha(x|y)\omega(y)\;,
\end{align*}
where $\tilde\mL_\alpha(x|y):=\alpha(x)\mL(y|x)/[\mL\alpha](y)$, according to the product rule of total probability. In other words, Jeffrey's rule promotes the inverse probability $\tilde\mL_\alpha(x|y)$ arising from Bayes theorem to a full-fledged \textit{channel} that propagates the agent's belief $\omega$ about the observation outcome back onto their belief $\alpha$ about the system's state. We stress again that the Bayesian inversion of $\mL$ is done with respect to the agent's prior $\alpha(x)$, while $\omega(y)$ can be completely arbitrary\footnote{As long as $\omega(y)>0 \implies [\mL\alpha](y)>0$, that is, as long as the new soft evidence does not falsify the entire stochastic model. To circumvent this problem, one can always assume that there is no such thing as ``absolute certainty'' and replace all zeros with arbitrarily small epsilons.}. It is then clear that, whenever $\omega(y)=\delta_{y,\bar y}$, that is, whenever the agent's observation has resulted in a definite outcome, then Jeffrey's rule reduces to Bayes' theorem. It has been later realized that Jeffrey's rule can in fact be derived from Bayes' theorem (plus some very natural assumptions) using Pearl's \textit{method of virtual evidence}~\cite{pearl, jaynes_2003,darwiche-chan-200567,jacobs-changing-mind}.

Recently, Jeffrey's rule has been discussed in relation with fluctuation relations and the second law of thermodynamics~\cite{buscemi-scarani-2021fluctuation,aw-buscemi-scarani}. The present work establishes another connection between Jeffrey's rule and statistical mechanics via the idea of observational entropy and coarse-grained states. For the sake of discussion, suppose for the time being that all POVM elements $\Pi_i$ corresponding to the coarse-graining $\C$ commute, so that they can all be diagonalized on the same orthonormal basis $\{\ket{x}\}_{x=1}^d$ as follows:
\begin{align*}
	\Pi_i=\sum_{x=1}^d p(i|x)\ketbra{x}\;.
\end{align*}
Obviously, the $\Pi_i$s form a POVM if and only if the numbers $p(i|x)$ form a conditional probability distribution. In other words, a POVM plays the role of the likelihood function in the example discussed above. In what follows we show that, whenever the $\Pi_i$'s commute, the coarse-grained state in Eq.~\eqref{eq:retrodiction} exactly coincides with the state of belief of an agent updated according to the Bayes--Jeffrey rule. To see this, begin by noticing that 
\begin{align*}
	V_i=\Tr\left[\Pi_i\right]=\sum_xp(i|x)\;.
\end{align*}
Therefore,
\begin{align*}
\sum_i\frac{p_i}{V_i}\;\Pi_i&=\sum_i\sum_x \frac{p_i}{\sum_{x'} p(i|x')} p(i|x)\ketbra{x}\\
&=\sum_ip_i\sum_x \frac{d^{-1}}{\sum_{x'} d^{-1}p(i|x')} p(i|x)\ketbra{x}\\
&\equiv\sum_ip_i\sum_x \frac{u(x)p(i|x)}{\sum_{x'} u(x')p(i|x')} \ketbra{x}\\
&=\sum_x\left(\sum_i \tilde p_u(x|i)p_i\right)\ketbra{x}\;,	
\end{align*}
where in the third line we introduced the dummy notation
$u(x)\equiv d^{-1}$ for the uniform probability and $\tilde p_u(x|i)=\frac{u(x)p(i|x)}{\sum_{x'} u(x')p(i|x')}$. The last line is in perfect agreement with Jeffrey's rule so that we can conclude that \emph{the coarse-grained state precisely corresponds to the agent's state of belief about the system, which, starting from a completely uninformative (i.e., uniform) prior, gets updated in the light of the new soft evidence provided by the outcome probability distribution} $p_i$.

At this point, since Theorem~\ref{thm:approx-lower-bound} holds for arbitrary (i.e., possibly non-commuting) POVMs, it is tempting to conclude that the coarse-grained state in  Eq.~\eqref{eq:retrodiction} represents a form of ``quantum retrodiction''. Such a conclusion would be in agreement with some previous works~\cite{barnett-pegg-jeffers,fuchs2002quantum,leifer-2006-q-dyn-as-cond-prob,Coecke-spekkens-2012-quantum-B-inference,leifer2013towards,PARZYGNAT-russo-2022-noncomm-Bayes-theorem}. However, it is fair to admit that, in the general non-commutative case, we don't have, strictly speaking, a generally accepted ``quantum Bayes theorem'' to appeal to: even the meaning of ``retrodiction'' in quantum theory is still debated~\cite{BJPsym21,surace-scandi-22-state-retrieval,axioms-retrodiction}.

We conclude this section with a comment about the difference between Bayesian retrodiction and linear inversion, which is what is done in tomographic reconstructions. When doing tomography, one treats the likelihood $\mL(y|x)$ as a matrix, representing a linear map $\mL$ from probability distributions on the index $x$ to probability distributions on the index $y$. Hence, in the ideal case, i.e., ignoring the problems that one encounters with finite statistics, it is natural to assume that the input to the reconstruction map is an element in the range of the linear map~$\mL$. In other words, it is natural to assume that the probability distribution on the data, based on which the reconstruction is done, correspond at least to \textit{some} input probability distribution via the linear map $\mL$. This is the case also if the linear map $\mL$ is invertible. On the contrary, in the case of Bayesian inference, nothing like that need to be assumed. This is true both for Jeffrey's rule and Bayes' theorem: indeed, it is almost never the case that delta distributions belong to the range of $\mL$, and yet Bayes' theorem is routinely applied also in such situations.

\section{Observational entropy in sequential measurements}
\label{sec:sequential}

Another key informational property of observational entropy is its monotonicity under sequential measurements~\cite{safranex2021generalized}. This property can be viewed as a consequence of the chain rule for classical relative entropy. Here we show how the basic property Theorem~\ref{thm:non-increase} (which appeared previously in~\cite{safranex2021generalized}) can also be considered in terms of Petz recovery, with a recovery based proof in Appendix~\ref{sec:app-proof-monotonicity-sequences}. Then we strengthen the previously known bound via explicit chain rule equalities.

\begin{theorem}\label{thm:non-increase} Consider two composable coarse-grainings $\C_1=\{\A_i\}_i$ and $\C_2=\{\B_j\}_j$ and their composition $\C_2\C_1=\{\B_j\circ\A_i\}_{ij}$, as depicted in Fig.~\ref{fig:coarse-graining}. Then we have 
\[\label{eq:decreasing-seq}
S_{\C_1}(\R)\ge S_{\C_2\C_1}(\R)\;,
\]
for any density matrix $\R$. The inequality becomes an equality if and only if (see Eq.~\eqref{eq:pi-vi-two})
\[\label{eq:ratios}
\frac{p_{ij}}{V_{ij}}=\frac{p_{i}}{V_{i}}\;,
\]
for all values $i$ and $j$. (In the above equation we assume, without loss of generality, that $\Pi_{ij}\neq 0$, namely, we remove events that never happen, so that $V_{ij}>0$ for all $i$ and $j$.)
\end{theorem}

\begin{remark}\label{rem:volume-channel}
Rephrasing the above theorem in more intuitive terms, the equality condition is reached when the first coarse-graining has disturbed the system's state so much that no new information can be acquired by the next measurement. This fact can be shown as follows. Notice that Eq.~\eqref{eq:ratios} can be rearranged as
\begin{align*}
    p_{ij}=\frac{V_{ij}}{V_{i}}p_{i}\;.
\end{align*}
Since $V_{ij}\ge 0$ and
\begin{align*}
d=\sum_{ij}V_{ij}=\sum_{i}V_{i}\;,
\end{align*}
one sees that the ratio $V_{ij}/V_{i}$ is in fact a normalized conditional probability of the index $j$ given the double index $(i,j)$. More explicitly, by denoting
\begin{align}\label{eq:conditional-volumes}
v_{j|i}:=\frac{V_{ij}}{V_{i}}\;,
\end{align}
we have that $v_{j|i}\ge 0$ and $\sum_{j}v_{j|i}=1$ for all values of index $i$. (Recall that we only have to consider possible events, i.e., those for which $\Pi_{ij}\neq 0$, which implies $V_{ij}=\tr{\Pi_{ij}}>0$ and also $V_i>0$, for all $i$ and $j$.)

Therefore, Theorem~\ref{thm:non-increase} simply says that the condition $S_{\C_2\C_1}(\R)= S_{\C_1}(\R)$ is equivalent to $p_{ij}=v_{j|i}p_{i}$. Namely, the propagation of the outcomes statistics from step $i$ to the next step $j$ happens in a memoryless fashion, as $v_{j|i}$ is independent of the initial state $\R$.
\end{remark}
\vspace{6pt}

Next, we generalize the above theorem by rewriting the difference as a classical relative entropy (Kullback-Leibler divergence) $\Dkl$ and further bounding it in terms of the quantum relative entropy.

\begin{framed}
\begin{theorem}[Generalization of Theorem~\ref{thm:non-increase}]
\label{thm:approx-non-increase}  Consider two composable coarse-grainings $\C_1=\{\A_i\}$ and $\C_2=\{\B_i\}$, and their composition $\C_2\C_1=\{\B_j\circ\A_i\}_{ij}$. Then,
\begin{align}\label{eqn:approx-sequential-theorem}
S_{\C_1}(\R)-S_{\C_2\C_1}(\R)= \Dkl\left(\bp\middle\|\bq\right)\;,
\end{align}
where the vector $\bp$ denotes the vector of joint probabilities $(\bp)_{ij}=p_{ij}$, while $\bq$ is the vector $(\bq)_{ij}=v_{j|i}p_{i}$,
for $v_{j|i}=V_{ij}/V_{i}$ the conditional volumes defined in~\ref{eq:conditional-volumes}.
\end{theorem}
\end{framed}

\begin{proof}
\begin{align*}
    S_{\C_1}(\R)-S_{\C_2\C_1}(\R)&=-\sum_{i} p_{i}\ln \frac{p_{i}}{V_{i}}+\sum_{i,j} p_{ij}\ln \frac{p_{ij}}{V_{ij}}=\sum_{i,j} p_{ij}\ln \frac{V_{i}p_{ij}}{p_{i}V_{ij}}=\Dkl\left(\bp\middle\|\bq\right)\;,
\end{align*}
where we have used $\sum_j p_{ij}=p_i$.
\end{proof}

\begin{remark}
The above theorem is a variation on the well-known chain rule for the relative entropy~\cite{CoverThomas}, $    \Dkl(p_{ij}\|q_{ij}) - \Dkl(p_i\|q_i) = \sum_i p_i \Dkl(p_{j|i}\|q_{j|i})
$. Theorem~\ref{thm:non-increase} follows as its corollary due to the non-negativity of classical relative entropy, from the zero-condition for relative entropy, and from the definitions of conditional probabilities.
\end{remark}
\bigskip

\begin{remark}
By definition of distributions $p_{i}=\tr{\A_i(\R)}$ and $p_{ij}=\tr{\B_j\circ\A_i(\R)}$, the first is the marginal distribution of the second, that is, $p_{i}=\sum_{j}p_{ij}$. Thus, in Theorem~\ref{thm:approx-non-increase}, $(\bp)_{ij}\equiv p_{ij}$ is the ``true'' joint probability distribution, while $(\bq)_{ij}\equiv v_{j|i} \, p_i$ is a distribution obtained by ignorantly propagating forward the marginal using only the volume terms.
\end{remark}
\bigskip  

By introducing the qc-channel corresponding to coarse-graining $\C_2$, defined in analogy with Eq.~\eqref{eq:M} as
\begin{align*}
\mN(\bullet):=\sum_j\tr{\B_j(\bullet)}\ketbra{j}\;,
\end{align*}
we see that the relation~\eqref{eqn:approx-sequential-theorem} can be rewritten as
\begin{align}
S_{\C_1}(\R)-S_{\C_2\C_1}(\R)&= \Dkl\left(\bp\middle\|\bq\right)\nonumber\\
&=\sum_{i}p_i\left\{\sum_jp_{j|i}\ln \frac{p_{j|i}}{v_{j|i}}\right\}\nonumber\\
&=\sum_i\tr{\A_i(\rho)} D(\mN(\R_i)\| \mN(u_i))\nonumber\\
&\le \sum_i\tr{\A_i(\rho)} D(\R_i\| u_i)\;, \label{eq:converse}
\end{align}
where $\R_i:=\A_i(\R)/\tr{\A_i(\R)}$ and $u_i:=\A_i(u)/\tr{\A_i(u)}$ are the states emerging out of the first coarse-graining, and where in~\eqref{eq:converse} we used the monotonicity of quantum relative entropy. This provides an upper bound on how much additional information can be extracted by performing additional sequential measurements. We formalize this result as a theorem.

\begin{framed}
\begin{theorem}[Maximal information gain]
\label{thm:sandwich}  Consider a coarse-graining $\C_1$. For any subsequent coarse-graining $\C_2$ and any state $\rho$, we have
\begin{align}\label{eq:upper_bound_ineq}
0\le S_{\C_1}(\R)-S_{\C_2\C_1}(\R)\le \sum_ip_iD(\rho_{i} \| u_{i})\;,
\end{align}
where $\R_i:=\A_i(\R)/\tr{\A_i(\R)}$ and $u_i:=\A_i(u)/\tr{\A_i(u)}$ are the post-measurement states corresponding to $\rho$ and the maximally mixed state $u$ (cf. Eq.~\eqref{eq:E_as_relativeE}), respectively.
\end{theorem}
\end{framed}

\begin{remark}
The above theorem provides a bound on the information that can be extracted sequentially. If the post-measurement states $\rho_{i}$ and $u_{i}$ are all equal, then no additional measurement can extract any further information. This is the case, for example, after a coarse-graining of the \textit{Gordon--Louisell} or \textit{measure-and-prepare} type~\cite{gordon-louisell,horodecki2003entanglement}, that is, a coarse-graining such that the post-measurement quantum state, given the classical outcome, is independent of the input state\footnote{Equivalently, a measure-and-prepare coarse-graning can be thought of as a device that measures a POVM and, dependening on the outcome obtained, prepares a fixed output state.}. Notice that the right-hand side of~\eqref{eq:upper_bound_ineq} does not depend on the second measurement, so its being non-zero value gives a possibility of the existence of such a measurement that can extract more. On the other hand, Theorem~\ref{thm:approx-non-increase} implies that if $\bp\neq \bq$, the additional measurement is guaranteed to provide extra information.
\end{remark}\bigskip 

\begin{remark}
The quantity $\sum_ip_iD(\rho_{i} \| u_{i})$ puts a bound on the maximum amount of information that can be still gained by performing an additional measurement. This also mean that it puts a bound on the amount of information that was irretrievably lost during the preceding measurements and cannot be recovered. Assuming that the second measurement $\C_2$ is the best informative measurement that we can perform, we define the  information lost due to $\C_1$ as
\[
I_{\C_1}^{\mathrm{lost}}(\R):=\inf_{\C_2}S_{\C_2\C_1}(\R)-S(\R).
\]
This measures the amount of information by which we fail to approach the lower bound given by the von Neumann entropy, despite our best efforts with the second measurement.
Rearranging the inequality in Eq.~\eqref{eq:upper_bound_ineq} we obtain a computable lower bound on the lost information as follows:
\begin{align}\label{eq:lost_inf_bound}
I_{\C_1}^{\mathrm{lost}}(\R)\ge S_{\C_1}(\R)-S(\R)-\sum_ip_iD(\rho_{i} \| u_{i})\;.
\end{align}	
Note however, that while $I_{\C_1}^{\mathrm{lost}}(\R)$ is non-negative by definition, the above lower bound may become negative, and thus trivial.
\end{remark}

\subsection{Achievability of von Neumann entropy by multiple sequential measurements}

Although formally similar, Theorems~\ref{th:lower-bound-vN-E} and~\ref{thm:non-increase} tell us about two different aspects of the observational entropy. Indeed, by combining them together, we obtain
\begin{align*}
S_{\C_1}(\rho)\ge S_{\C_2\C_1}(\rho)\ge S(\rho)\;,
\end{align*}
which straightforwardly extends to longer sequences as
\begin{align*}
   S_{\C_1}(\rho)\ge S_{\C_2\C_1}(\rho)\ge  \cdots \ge S_{\C_n\cdots\C_2\C_1}(\rho) \ge \cdots\ge S(\rho)\;, 
\end{align*}
that is, the more measurements are performed, the closer to the ultimate lower bound $S(\rho)$---that is, the closer to the state of maximum knowledge about $\rho$---one can get. There is a limit, however: with each step, the post-measurement states resulting from $\rho$ converge closer to those obtained from a maximally mixed state, meaning that also less and less information about $\rho$ can be extracted at each step. When the extractable information becomes zero, that is exactly the point at which the equality in Eq.~\eqref{eq:decreasing-seq} holds. Such a balanced relation between information extracted and disturbance caused by the measurements was already discussed in Refs.~\cite{safranek2019b,safranex2021generalized}. Theorem~\ref{thm:sandwich} makes this relation significantly more explicit, through the relation with the lost information,
\[
S_{\C_1}(\rho)\ge  \cdots \ge S_{\C_n\cdots\C_2\C_1}(\rho) \ge I_{\C_1}^{\mathrm{lost}}(\R)+S(\R)\ge S(\rho)\;, 
\]
namely, the unattainability gap is fixed already by the first measurement, due to lost information being non-negative. A computable bound on this is in Eq.~\eqref{eq:lost_inf_bound}.

\section{Observational entropy and refinements of coarse-graining}
\label{sec:refine}

Having considered measurements in sequence, we now turn to the question of when one coarse-graining is finer than another. A basic and intuitive fact is that observational entropy is monotonically non-decreasing under stochastic post-processings of the outcomes~\cite{martens1990nonideal,buscemi2005clean}. That is, if the outcomes statistics of one coarse-graining $\C'$ can be recovered from those of another coarse-graining $\C$, then the statistics of $\C$ is \emph{sufficient} for that of $\C'$. Intuitively speaking, $\C$ yields more information (and thus a smaller observational entropy) than $\C'$.

We begin with the formal definition of refinements (extending that of a ``finer vector of coarse-grainings''~\cite{safranek2019b,safranex2021generalized}).

\begin{definition}[Refinements]\label{def:finer_set_coarse_graining}
	We say that a coarse-graining $\C=\{\A_i\}_i$ is a \textit{refinement} of another coarse-graining $\C'=\{\A'_j\}_j$ (or alternatively, that $\C'$ is a \textit{post-processing} of $\C$), and denote this by
	\[\nonumber
	{\C}\hookleftarrow {\C'}\;,
	\]
	whenever there exists a stochastic matrix $t$, $t_{j|i}\ge 0$, $\sum_{j} t_{j|i}=1$ for all $i$, such that the corresponding POVM elements satisfy the relation
\[
\quad \Pi'_{j}=\sum_{i}t_{j|i}\Pi_i\;,\quad\forall {j}\;.
\]
\end{definition}
\noindent
Notice that the indices $i$ and $j$ in the above definition could as well be taken to be multi-indices. 

The following theorem is yet another direct consequence of the monotonicity of the relative entropy, and generalizes Theorem 2 in~\cite{safranex2021generalized}. Continuing the theme, we provide in Appendix~\ref{app:monotonic} a proof based on the Petz theorem.

\begin{theorem}[Stochastic Monotone]\label{thm:monotonic} Observational entropy is a monotonic function of the ``coarseness'' of coarse-graining. More precisely, given two coarse-grainings $\C$ and $\C'$, respectively associated with POVMs $\{\Pi_i\}_i$ and $\{\Pi'_j\}_j$, if ${\C}\hookleftarrow {\C'}$ as in Definition~\ref{def:finer_set_coarse_graining}, then
\[
S_{\C}(\R)\le S_{\C'}(\R)\;,
\]
for any state $\R$. The inequality becomes equality if and only if, for all $i$,
\[\label{eq:equality-bayes}
\frac{p_i}{V_i}=\sum_{j} t_{j|i}\frac{p'_{j}}{V'_{j}}\;,
\]
where $p'_{j}:=\Tr[\Pi'_{j}\, \rho]$ and $V'_{j}:=\Tr[\Pi'_{j}]$.
\end{theorem}

\begin{remark}\label{rem:quasi-B-inversion}
    
    The condition for equality, Eq.~\eqref{eq:equality-bayes}, gets a very clear interpretation once we realize that the ratio $t_{j|i}V_i/V'_j$ is itself a normalized conditional probability of $i$ given $j$. This is a consequence of the fact that $V'_j$ at the denominator equals by definition $\sum_i t_{j|i}V_i$. Therefore, defining
    \begin{align*}
        \tilde t_{i|j}:=\frac{t_{j|i}V_i}{V'_j}\;,
    \end{align*}
    the condition for equality becomes $p_i=\sum_j\tilde t_{i|j}p'_j$, that is, not only the measurement statistics obtained from $\C$ is sufficient for $\C'$ (as it was assumed from the beginning), but also the vice versa holds---i.e., the two coarse-grainings $\C$ and $\C'$ provide equivalent measurement statistics.
\end{remark}\bigskip 

The refinement theorem can also be strengthened in the form of an identity. 

\begin{framed}
\begin{theorem}[Generalization of Theorem~\ref{thm:monotonic}]
\label{thm:approx-monotonic2}
Given two coarse-grainings $\C$ with POVM elements $\{\Pi_i\}$ and $\C'$ with POVM elements $\{\Pi'_j\}$, suppose that ${\C}\hookleftarrow {\C'}$ (see Def.~\ref{def:finer_set_coarse_graining}). Then we have the following equality and inequality,
\begin{align}
S_{\C'}(\R)-S_{\C}(\R)&= \Dkl(\bp_2\|\bq_2),\\
&\ge \Dkl(\bp\|\bq),
\end{align}
where
\[
    (\bp_2)_{ij}=t_{j|i}p_i,\quad\quad (\bq_2)_{ij}=\tilde{t}_{i|j}p_j',
\]
where we have defined backward stochastic element $\tilde{t}_{i|j}={t_{j|i}V_i}/{V_j'}$. Their marginals are
\[
    (\bp)_{i}=p_i,\quad\quad (\bq)_{i}=\sum_{j}\tilde{t}_{i|j} p_j'.
\]
The following statements are equivalent: $S_{\C'}(\R)-S_{\C}(\R)=0$ iff $\bp_2=\bq_2$ iff $\bp=\bq$.
\end{theorem}
\end{framed}

\begin{proof} \emph{Equality:}
Using $\sum_j t_{j|i}=1$ and $p_j'=\sum_i t_{j|i}p_i$ we have
\begin{align*}
    S_{\C'}(\R)-S_{\C}(\R)&=-\sum_{j} p_{j}'\ln \frac{p_{j}'}{V_{j}'}+\sum_{i} p_{i}\ln \frac{p_{i}}{V_{i}}\\
    &=-\sum_{i,j}t_{j|i} p_i\ln \frac{p_{j}'}{V_{j}'}+\sum_{i,j} t_{j|i}p_{i}\ln \frac{p_{i}}{V_{i}}\\
    &=\sum_{i,j} t_{j|i}p_{i}\ln \frac{t_{j|i}p_{i}}{\frac{t_{j|i}V_{i}}{V_{j}'}p_{j}'}\\
    &=\Dkl\left(\bp_2\middle\|\bq_2\right).
\end{align*}
\emph{Inequality:} Follows from monotonicity of the relative entropy, which shows that relative entropy of a joint distribution is larger than that of its marginals.\\
\emph{Equality-to-zero conditions:} Follow from the zero-condition for relative entropy and from Theorem~\ref{thm:monotonic}.
\end{proof}

\section{Concavity properties of observational entropy}
\label{sec:concave}

Geometric properties of the observational entropy can also be seen as consequences of the relative entropy monotonicity. Here we exhibit concavity as both a function of states $\rho$ and coarse-grainings $\C$. These properties derive from joint convexity of the classical relative entropy. Here we relate them to the present discussion with a proof based on monotonicity in Appendix~\ref{sec:app-proof-concavity}.

\begin{theorem}\label{thm:concavity}
The observational entropy $S_{\C}(\rho)$ is concave both in the coarse-graining $\C$ and the state $\rho$. That is:
\begin{enumerate}
    \item state concavity: let $\{\R_k\}$ be a family of density matrices and $\lambda_k$ a probability distribution; then
    \begin{align}
    \sum_k\lambda_k S_\C(\R_k)\le S_\C\left(\sum_k\lambda_k\R_k\right)\;,
    \end{align}
    with equality if and only if all the $\R_k$'s are indistinguishable for $\C$, that is, they induce the same probabilities $\tr{\Pi_i(\R_k)}=\tr{\Pi_i(\R_{k'})}$ for all $i,k,k'$.
    
    \item coarse-graining concavity: let $\{\C_k\}$ be a family of coarse-grainings, 
    with corresponding POVMs $\{\Pi_{i|k}\}_i$, and $\lambda_k$ a probability distribution; putting $\C:=\sum_k\lambda_k\C_k$ and correspondingly $\Pi_i:=\sum_k\lambda_k\Pi_{i|k}$, we have
    \[\label{eqn:52}
    \sum_k\lambda_k S_{\C_k}(\R)\le S_{\C}(\R),
    \]
    with equality iff\begin{align}\label{eqn:concavity-equality}
    p_{i|k}= \frac{p_{i}}{V_{i}}V_{i|k}
    \end{align}
for all $i$ and $k$. Here, we defined
\[
p_{i|k}=\Tr[\Pi_{i|k}\ \rho],\quad V_{i|k}=\Tr[\Pi_{i|k}],\quad p_{i}=\Tr[\Pi_{i}\ \rho],\quad V_{i}=\Tr[\Pi_{i}].
\]
\end{enumerate}
\end{theorem}

\begin{remark}
    \label{rem:concavity-equality}
    Whenever $\Pi_{i|k}$ is nonzero, the equality condition \eqref{eqn:concavity-equality} implies that each of the measurements (labelled by $k$) induces the same probability to volume ratio,
    \begin{align}
        \frac{p_{i|k}}{V_{i|k}} = \frac{p_{i}}{V_{i}} \;.
    \end{align}
    For POVMs with completely disjoint outcome sets (that is, only one $\Pi_{i|k}$ nonzero for each $i$), equality in \eqref{eqn:52}, and therefore in~\eqref{eqn:concavity-equality}, always holds.
\end{remark}

\section{Conclusion}

In this paper we demonstrated how to describe observational entropy as a quantum relative entropy, and---focusing on the relative entropy monotonicity under CPTP maps and the associated theory of approximate recoverability---analyzed observational entropy from an information-theoretic perspective. 

This approach helped to clarify the relationship between coarse-grained measurements, entropies, and states. In particular we showed that the state obtained from Petz recovery (on the quantum-classical measurement channel implementing observational entropy) takes the simple form (cf.~Theorems~\ref{th:lower-bound-vN-E} and~\ref{thm:approx-lower-bound}), 
\begin{equation}\label{eq:final_quantumBayes}
    \Rrec = \sum_i p_i \, \Pi_i / V_i\;,
\end{equation}
and is also the state retrodicted experimentally by Bayesian analysis of measurement outcomes. It provides a representation of the observer's state of knowledge, which, starting from a completely uninformative (i.e., uniform) prior, gets updated in the light of the new soft evidence provided by the outcome probability distribution $p_i$. Theorem~\ref{thm:approx-lower-bound} shows that the accuracy, measured by quantum relative entropy, quantum fidelity, and trace distance, of the description about the ``true'' state provided by such a coarse-grained representation can be very well quantified in terms of the difference between the observational entropy and the von Neumann entropy, which thus gains a new operational meaning.

Viewed one way---considering particular coarse-grainings related to thermodynamics---these lower bounds can provide a useful tool in statistical and thermodynamic analysis. Meanwhile, from the information-theoretic perspective they may have even greater value, providing accessible upper bounds on state estimation error, and thereby affording a potential route towards confident state retrodiction as an alternative to tomography.

Given that observational entropy is interpreted as a measure of the observers' uncertainty about a quantum system, according to Theorem~\ref{thm:approx-lower-bound}, reducing this uncertainty also means reducing the difference between the true and the coarse-grained state. Thus, this inequality provides supporting evidence for formalizing the updating rule of Eq.~\eqref{eq:final_quantumBayes} as the ``quantum Bayes theorem'', adding to the ongoing discussion on the meaning of ``retrodiction'' in quantum theory~\cite{BJPsym21,surace-scandi-22-state-retrieval,axioms-retrodiction}.

The framework of observational entropy appears to be, in this work and elsewhere, a fertile and physically motivated playground for relating topics throughout quantum information theory, measurement theory, and statistical thermodynamics. It has already revealed a number of interesting connections, with natural applications in open and isolated non-equilibrium thermodynamics, quantum correlations and entanglement, and, as seen here, state retrodiction and sequential measurement theory. Many more such connections certainly remain to be studied.

\section*{Acknowledgments}
F.~B. acknowledges support from MEXT Quantum Leap Flagship Program (MEXT QLEAP) Grant No. JPMXS0120319794; from MEXT-JSPS Grant-in-Aid for Transformative Research Areas (A) ``Extreme Universe'', No. 21H05183; from JSPS KAKENHI Grants No.~20K03746 and No.~23K03230. D.~\v{S}. acknowledges support from the Institute for Basic Science in South Korea IBS-R024-D1. J.~S. acknowledges support by by MICIIN with funding from European Union NextGenerationEU (PRTR-C17.I1) and by Generalitat de Catalunya.


\bibliographystyle{unsrturl}
\bibliography{library,ads}

\appendix

\section{Proof of Theorem~\ref{th:lower-bound-vN-E}}
\label{sec:app-proof-of-bound-with-vonN-entropy}

The inequality follows directly from Theorem~\ref{thm:inequality_for_channels} applied to Eq.~\eqref{eq:diff-ent-diff-rel}. To study the equality condition, consider that in this case we have $\S=\openone/d$ and 
\begin{align}
	\mM(\bullet)&=\sum_i\tr{\Pi_i\ \bullet}\ketbra{i}\\
	&=\sum_i\tr{\A_i(\bullet)}\ketbra{i}\\
	&=\sum_i\tr{\sum_m K_{i m}\bullet K_{i m}^\dag}\ketbra{i}\label{eq:choice-of-kraus}\\
	&=\sum_{i,m,k}\ket{i}\bra{k}K_{i m}\bullet K_{i m}^\dag\ket{k}\bra{i}\;,
\end{align}
where we used a Kraus decomposition $\A_i(\bullet)=\sum_m K_{im}\bullet K_{im}^\dag$. In particular,
\begin{align*}
	\mM(\openone)=\sum_i V_i\ketbra{i}\;,
\end{align*}
so that the Petz recovery map~\eqref{eq:Petz_map} for $\mM$ computed with respect to the uniform reference state $\openone/d$ is
\begin{align}
	\mMS(\bullet)&:=\mM^\dag(\mM(\openone)^{-1/2}\bullet\mM(\openone)^{-1/2})\nonumber\\
	&=\sum_{i,m,k}\frac{K_{i m}^\dag\ket{k}\bra{i}\bullet\ket{i}\bra{k}K_{i m}}{V_i}\label{eq:tildeMid}\\
	&=\sum_{i}\frac{\Pi_i}{V_i}\bra{i}\bullet\ket{i}\;,\nonumber\label{eq:petz-transpose}
\end{align}
where in the last line we used the facts that the POVM element $\Pi_i$ corresponding to the map $\A_i$ equals $\sum_m K_{i m}^\dag K_{i m}$, and the identity $\sum_k\ketbra{k}=\openone$.
The equality condition is then
\begin{align}
	\R&=\mMS(\mM(\R))\\
	&=\mMS\left(\sum_{i}\frac{\tr{\A_{i}(\R)}}{V_i}\ketbra{i}\right)\\
	&=\sum_{i}\frac{p_i}{V_i}\Pi_i=\Rrec\;,
\end{align}
as claimed. Notice how the choice of a particular Kraus representation in~\eqref{eq:choice-of-kraus} is immaterial for the argument.

\section{Proof of Theorem~\ref{thm:non-increase}}
\label{sec:app-proof-monotonicity-sequences}

	Let $\A_i$ be the CP maps associated with the initial coarse-graining $\C_1=\{\A_i\}$, and let $\A_{ij}$ be the CP-maps associated with the extended coarse-graining $\C_2\C_1=\{\A_{ij}\}$, that is, $\A_{ij}(\bullet):=\B_{j}(\A_{i}(\bullet))$. Moreover, following Eq.~\eqref{eq:M}, let us introduce the following CPTP maps:
\[
\mM^2(\bullet)=\sum_{i,j}\tr{\A_{ij}(\bullet)}\ketbra{i,j}\;,
\]
and
\[\label{eq:Mn}
\mM(\bullet)=\sum_i\tr{\A_{i}(\bullet)}\ketbra{i}\;.
\]
Defining the CPTP map $\Tr_2$ as the partial trace over the second index, we have
\[
\mM=\Tr_2\circ\mM^2\;.
\]
Then, according to Eq.~\eqref{eq:obsE_as_relativeE}, we obtain
\begin{align}
	S_{\C_2\C_1}&=\ln d-D(\mM^2(\R)\|\mM^2(\openone/d))\\
	&\le \ln d-D(\Tr_2(\mM^2(\R))\|\Tr_2(\mM^2(\openone/d)))\label{eq:inequality}\\
	&=\ln d-D(\mM(\R)\|\mM(\openone/d))\label{eq:obsE_from_Mn}\\
	& = S_{\C_1}\;.
\end{align}

Let us now look into the condition for equality. Defining $\tilde{\R}:=\mM^2(\R)$ and $\tilde{\S}:=\mM^2(\openone/d)$, according to Theorem~\ref{thm:inequality_for_channels} we have the equal sign in~\eqref{eq:inequality} if and only if the following CPTP map
\[\label{eq:recoverymap}
\tilde{\mE}_{\tilde\S}(\bullet):=\sqrt{\tilde\S}\Tr_2^\dag\left[\frac{1}{\sqrt{\Tr_2(\tilde\S)}}\bullet\frac{1}{\sqrt{\Tr_2(\tilde\S)}}\right]\sqrt{\tilde\S}
\]
satisfies
\[\label{eq:eqcondition}
\tilde{\mE}_{\tilde\S}(\Tr_2(\tilde \R))=\tilde \R\;.
\]
By explicit computation,
\begin{align*}
\tilde \S&=\sum_{i,j} \frac{V_{ij}}{d}\ketbra{i,j}\;,\\
 \Tr_2(\tilde \S)&=\sum_{i} \frac{V_{i}}{d}\ketbra{i}\;,\\ \Tr_2^\dag&=\sum_{j}\ket{j}\bullet\bra{j}\;,
\end{align*}
which gives
\[
\tilde{\mE}_{\tilde\S}(\bullet)=\sum_{i,\tilde i,j}\sqrt{\frac{V_{ij}}{V_{i}}}\sqrt{\frac{V_{\tilde ij}}{V_{\tilde i}}}\ket{i,j}\bra{i}\bullet\ket{\tilde i}\bra{\tilde i,j}\;.
\]
Considering that
\begin{align*}
\tilde \R&=\sum_{i,j}p_{ij}\ketbra{i,j}\;,\\
\Tr_2(\tilde \R)&=\sum_{i}p_{i}\ketbra{i}\;,
\end{align*}
the inequality~\eqref{eq:inequality} becomes identity when
\[
\sum_{i,j}\frac{V_{ij}}{V_{i}}p_i\ket{i,j}\bra{i,j}=\sum_{i,j}p_{ij}\ketbra{i,j}\;,
\]
i.e., when for all values of $i,j$,
\[
p_{ij}=\frac{V_{ij}}{V_{i}}p_{i}\;,
\]
as claimed.

\section{Proof of Theorem~\ref{thm:monotonic}}\label{app:monotonic}

A proof of Theorem~\ref{thm:monotonic} follows straightforwardly from monotonicity of classical relative entropy under classical stochastic channels. However, below we give a slightly longer proof leveraging the quantum monotonicity, in the spirit of presenting a unified analysis. 

\begin{proof}
	
Using the definition of $\mM$ as in Eq.~\eqref{eq:Mn}, and the corresponding expression for observational entropy $S_{\C}$, Eq.~\eqref{eq:obsE_from_Mn}, we define correspondingly $\mM'$ for $S_{\C'}$:
\begin{align*}
	\mM'(\bullet)&:=\sum_{j}\tr{\Pi'_{{j}}\ \bullet}\ketbra{{j}}\\
	&=\sum_{j}\tr{\sum_{i}t_{j|i}\Pi_i\ \bullet}\ketbra{{j}}\\
	&=\sum_{i,{j}} t_{j|i}\tr{\Pi_i\ \bullet}\ketbra{{j}}\\
	&=\D\left(\sum_{i}\tr{\Pi_i\ \bullet}\ketbra{i}\right)\\
	&=(\D\circ \mM)(\bullet)\;,
\end{align*}
where we denoted by $\D$ the channel
\[
\D(\bullet):=\sum_{i,{j}}t_{j|i}\ket{{j}}\bra{i}\bullet \ket{i}\bra{{j}}\;.
\]
From Theorem~\ref{thm:inequality_for_channels}, applied to $\tilde{\R}:=\mM(\R)$ and $\tilde{\S}:=\mM(\openone/d)$, the equality condition $S_{\C}(\R)= S_{\C'}(\R)$ becomes $D(\tilde{\R}\|\tilde{\S})=D(\D(\tilde{\R})\|\D(\tilde{\S}))$. Hence, by specializing Eqs.~\eqref{eq:recoverymap} and~\eqref{eq:eqcondition} to the present case, we have
\begin{align*}
	\D^\dag(\bullet)&=\sum_{i,{j}}t_{j|i}\ket{i}\bra{{j}}\bullet \ket{{j}}\bra{i}\;,\\
	 \tilde \R&=\sum_i p_i\ketbra{i}\;,\\
	  \tilde \S&=\sum_i\frac{V_i}{d}\ketbra{i}\;,\\
	\D(\tilde \R)&=\sum_{j} p'_{j}\ketbra{{j}}\;,\\
	\D(\tilde \S)&=\sum_{j}\frac{V'_{j}}{d}\ketbra{{j}}\;,   
\end{align*}
where we have used $p'_{j}=\sum_{i} t_{j|i}p_{i}$ and $V'_{j}=\sum_{i} t_{j|i}V_{i}$. 
This gives
\begin{align}
\tilde{\D}_{\tilde\S}(\bullet)&:=\sqrt{\tilde\S}\;\D^\dag\left[\frac{1}{\sqrt{\D(\tilde\S)}}\bullet\frac{1}{\sqrt{\D(\tilde\S)}}\right]\sqrt{\tilde\S}\\
&=\sum_{i,{j}}\frac{ t_{j|i}V_i}{V'_{j}}\ket{i}\bra{j}\bullet\ket{j}\bra{i}.
\end{align}
The inequality becomes identity, i.e., $S_{\C}(\R)= S_{\C'}(\R)$, if and only if $\tilde{\D}_{\tilde\S}(\D(\tilde \R))=\tilde \R$, which we can rewrite as
\[
\sum_{i,{j}}\frac{ t_{j|i}V_i}{V'_{j}}p'_{j}\ket{i}\bra{i}=\sum_i p_i\ketbra{i}\;,
\]
that is, if and only if
\[
p_i=\sum_{j} \frac{t_{j|i}V_i}{V'_{j}}p_{j}'\;,
\]
for all $i$, as claimed.

\end{proof}

\section{Proof of Theorem~\ref{thm:concavity}}
\label{sec:app-proof-concavity}

An alternative proof can be given using the joint convexity of classical relative entropy. However, the one given below, which is based on Petz's recovery theorem, provides additional insight into the current framework.

\begin{proof}

Both state concavity and coarse-graining concavity follow from Eq.~\eqref{eq:diff-ent-diff-rel} and Theorem~\ref{thm:inequality_for_channels}.

To prove the state concavity property, let $\rho=\sum_k\lambda_k\rho_k$ and define an extended (block-form) state
\begin{align*}
	\overline{\R}:=\sum_k\lambda_k\rho_k\otimes\ketbra{k}_E\;.
\end{align*}
Clearly, taking the partial trace over the extension $E$ gives us back $\R$, i.e., $\Tr_E[\overline{\R}]=\rho$, while we denote the other marginal as $\rho_E:=\sum_k\lambda_k\ketbra{k}_E$. Consider now that
\begin{align}
	&D\Big((\mM\otimes\id_E)(\overline{\R})\Big\|\mM(\openone/d)\otimes\rho_E\Big)\label{eq:blocks}\\
	&= D\left(\sum_k\lambda_k\mM(\rho_k)\otimes\ketbra{k}_E\left\|\sum_k\lambda_k\mM(\openone/d)\otimes\ketbra{k}_E\right.\right)\nonumber\\
	&=\sum_k\Tr\left[ \lambda_k\mM(\rho_k)\ \Big\{\log(\lambda_k\mM(\rho_k))-\log(\lambda_k\mM(\openone/d) )\Big\}\right]\nonumber\\
	&=\sum_k\Tr\left[ \lambda_k\mM(\rho_k)\ \Big\{\log(\mM(\rho_k))-\log(\mM(\openone/d) )\Big\}\right]\nonumber\\
	&=\sum_k\lambda_k D(\mM(\rho_k)\|\mM(\openone/d))\nonumber\\
	&=\log d-\sum_k\lambda_k S_\C(\R_k)\;,\nonumber
\end{align}
and therefore,
\begin{align*}
	\sum_k\lambda_k S_\C(\R_k)&=\log d - D\Big((\mM\otimes\id_E)(\overline{\R})\Big\|\mM(\openone/d)\otimes\rho_E\Big)\\
	&\le \log d - D\Big(\Tr_E[(\mM\otimes\id_E)(\overline{\R})]\Big\|\Tr_E[\mM(\openone/d)\otimes\rho_E]\Big)\\
	&= \log d - D\Big(\mM(\rho)\Big\|\mM(\openone/d)\Big)\\
	&=S_\C(\rho)\;,
\end{align*}
where the inequality comes again from Theorem~\ref{thm:inequality_for_channels}. The same theorem gives us also the condition for equality, that is
\begin{align*}
	&(\mM\otimes\id_E)(\overline{\R})\\
	&=\sqrt{\mM(\openone/d)\otimes\rho_E}\Tr_E^\dag \left[\frac{1}{\sqrt{\mM(\openone/d)}}\mM(\Tr_E[\overline{\R}])\frac{1}{\sqrt{\mM(\openone/d)}}\right] \sqrt{\mM(\openone/d)\otimes\rho_E}\\
	&=\sum_{i,k}\lambda_k p(i|k)\ketbra{i}\otimes\rho_E\;,
\end{align*}
where $\Tr_E^\dag(\bullet):=\bullet\otimes\openone_E$ and $p(i|k):=\Tr[\Pi_i\ \rho_k]$. The above can be rewritten as
\begin{align*}
	\sum_{k}\lambda_k\left(\sum_ip(i|k)\ketbra{i}\right)\otimes\ketbra{k}_E=\sum_k\lambda_k\left[\sum_{i}\left(\sum_{k'}\lambda_{k'}p(i|k')\right)\ketbra{i}\right]\otimes\ketbra{k}_E\;,
\end{align*}
and since $\lambda_k>0$, this is possible if and only if $p(i|k)=p(i|k')$ for all $i,k,k'$, as claimed.

To prove the coarse-graining concavity property, we can proceed along the same lines, this time constructing an extended CPTP map as follows:
\begin{align*}
	\overline{\mM}(\bullet):=\sum_k\lambda_k\mM_k(\bullet)\otimes\ketbra{k}_E\;,
\end{align*}
where $\mM_k$'s are the CPTP maps corresponding to the coarse-grainings $\C_k$. In this case, proceeding along exactly the same lines as in~\eqref{eq:blocks}, we have
\begin{align*}
	&D\Big(\overline{\mM}(\R)\Big\|\overline{\mM}(\openone/d)\Big)\\
	&= D\left(\sum_k\lambda_k\mM_k(\rho)\otimes\ketbra{k}_E\left\|\sum_k\lambda_k\mM_k(\openone/d)\otimes\ketbra{k}_E\right.\right)\nonumber\\
	&=\log d - \sum_k\lambda_kD(\mM_k(\rho)\|\mM_k(\openone/d))\\
	&=\log d- \sum_k\lambda_k S_{\C_k}(\R)\;.
\end{align*}
Therefore
\begin{align*}
	\sum_k\lambda_k S_{\C_k}(\R)&=\log d-D\Big(\overline{\mM}(\R)\Big\|\overline{\mM}(\openone/d)\Big)\\
	&\le \log d-D\Big(\Tr_E\circ\overline{\mM}(\R)\Big\|\Tr_E\circ\overline{\mM}(\openone/d)\Big)\\
	&=\log d - D\left(\left(\sum_k\lambda_k\mM_k\right)(\R)\left\|\left(\sum_k\lambda_k\mM_k\right)(\openone/d)\right.\right)\\
	&=S_\C(\R)\;,
\end{align*}
as claimed. Moving on to the equality condition, that is
\begin{align*}
    D\Big(\overline{\mM}(\R)\Big\|\overline{\mM}(\openone/d)\Big)=D\Big(\Tr_E\circ\overline{\mM}(\R)\Big\|\Tr_E\circ\overline{\mM}(\openone/d)\Big)\;,
\end{align*}
again, as a consequence of Theorem~\ref{thm:inequality_for_channels}, we know that the above holds if and only if
\begin{align*}
	&\overline{\mM}(\R)\\
	&=\sqrt{\overline{\mM}(\openone/d)}\Tr_E^\dag \left\{\frac{1}{\sqrt{\Tr_E\left[\overline{\mM}(\openone/d)\right]}}\Tr_E\left[\overline{\mM}(\R)\right]\frac{1}{\sqrt{\Tr_E\left[\overline{\mM}(\openone/d)\right]}}\right\} \sqrt{\overline{\mM}(\openone/d)}\\
	&=\sqrt{\overline{\mM}(\openone)}\Tr_E^\dag \left\{\frac{1}{\sqrt{\Tr_E\left[\overline{\mM}(\openone)\right]}}\Tr_E\left[\overline{\mM}(\R)\right]\frac{1}{\sqrt{\Tr_E\left[\overline{\mM}(\openone)\right]}}\right\} \sqrt{\overline{\mM}(\openone)}\\
	&=\sqrt{\overline{\mM}(\openone)}\Tr_E^\dag \left\{\sum_i \frac{\sum_{k}\lambda_{k}p_{i|k}}{\sqrt{\sum_{k'}\lambda_{k'}V_{i|k'}}\sqrt{\sum_{k''}\lambda_{k''}V_{i|k''}}}\ketbra{i}\right\} \sqrt{\overline{\mM}(\openone)}\\
	&=\sqrt{\overline{\mM}(\openone)}\left\{\sum_i \frac{\sum_{k}\lambda_{k}p_{i|k}}{\sum_{k'}\lambda_{k'}V_{i|k'}}\ketbra{i}\otimes\openone_E\right\} \sqrt{\overline{\mM}(\openone)}\\
	&=\sum_{k''}\sum_i \lambda_{k''}V_{i|k''} \frac{p_i}{V_i}\ketbra{i}\otimes\ketbra{k''}_E\;,
\end{align*}
where in the last line we used the identity $p_i=\Tr[\Pi_i\rho]=\Tr[\sum_k\lambda_k\Pi_{i|k}\rho]=\sum_k\lambda_kp_{i|k}$, and analogously for $V_i=\sum_k\lambda_kV_{i|k}$.

Summarizing, we showed that $\sum_k\lambda_k S_{\C_k}(\R)=S_\C(\R)$ if and only if
\begin{align*}
    \sum_{k''}\sum_i \lambda_{k''}p_{i|k''}\ketbra{i}\otimes\ketbra{k''}_E=
\sum_{k''}\sum_i \lambda_{k''}V_{i|k''} \frac{p_i}{V_i}\ketbra{i}\otimes\ketbra{k''}_E\;,
\end{align*}
which in turns holds if and only if
\begin{align*}
    p_{i|k''} = V_{i|k''} \, \frac{p_i}{V_i}
\end{align*}
for all $i$ and all $k''$, i.e., the probability-to-volume ratios do not depend on the mixing index~$k''$. 
\end{proof}

\end{document}